\documentclass[11pt]{article}

\usepackage{amsmath,amsthm,amssymb,amscd,bbm,url}
\usepackage{authblk}
\usepackage{tikz}
\usepackage{subcaption}
\usepackage{booktabs}
\usepackage{xurl}
\usepackage{csquotes}
\usepackage{hyperref}
\usepackage[mathlines]{lineno}
\usepackage{natbib}

\usepackage[utf8]{inputenc}
\usepackage{geometry}
\usepackage{sectsty}
\usepackage[normalem]{ulem}

\sectionfont{\sffamily\bfseries\upshape\large}
\subsectionfont{\sffamily\bfseries\upshape\normalsize}
\subsubsectionfont{\sffamily\mdseries\upshape\normalsize}
\makeatletter
\renewcommand\@seccntformat[1]{\csname the#1\endcsname.\quad}

\makeatletter
\def\@maketitle{%
  \begin{center}%
  \let \footnote \thanks
    {\large \@title \par}%
    {\normalsize
      \begin{tabular}[t]{c}%
        \@author
      \end{tabular}\par}%
    {\small \@date}%
  \end{center}%
}
\makeatother

\begin{document}
\title{\bf A Bayesian analysis of home advantage in professional squash\footnote{Data for this analysis was provided by squashinfo.com and includes PSA World Tour matches from December 2018
to March 2024.}}
 \author{Philip Greengard\thanks{Work completed while at Columbia University, currently at Flatiron Institute, pgreenga@gmail.com}, Samer Takriti}

\date{10 June 2025}

\maketitle

\begin{abstract}
We estimate the effect of playing in one's home country in professional squash using a Bayesian hierarchical model applied to men's and women's Professional Squash Association matches from 2018-2024. The model incorporates players' world rankings and whether they are competing in their home country. Using margin of victory in games as our outcome, we estimate that home advantage adds 0.4 games for men and 0.3 games for women to the expected margin, with standard errors of 0.1. For evenly matched players, this effect corresponds to an increase in win probability from 50\% to roughly 58\% for men and 56\% for women. We estimate particularly strong home effects in Egypt, where many major tournaments are held, though data limitations prevent precise estimation of country-specific effects in many other nations.
\end{abstract}

\section{Introduction}
Professional Squash Association (PSA) tournaments are held in countries 
spanning the globe including New Zealand, Hong Kong, Malaysia, Qatar, Egypt, France, England, and the U.S. Each of these countries has its own squash community and 
when players play in their home country, the crowd usually gives its full support 
to the local player. Various forms of home court advantage has been studied for decades across a range of sports and contexts \citet{fischer2021, bilalic2021, 
gomez2011}.
In many cases, the effect of home advantage is significant \citet{jamieson2010, pollard2014}. 
Here we test the effect of playing in your home country on the PSA World Tour. 
We measure this effect with men's and women's matches from December 2018 until 
early 2024. 

We use a Bayesian hierarchical model \citet{gelman2006, gelman2013} to predict the outcome of matches that uses 
the world ranking of each player and whether one of the two players has the home 
advantage. 
We say that a player has home advantage if he/she is playing in his/her home country (as 
listed by squashinfo.com) and the other player is not. In all other cases, there is
no home advantage.  
Our model is fit to the margin of victory in games of each match. Each best of five 
match has an outcome of one of $-3, -2, -1, 1, 2, 3$. By using margin of victory as the
observation for the model
instead of the binary win/loss, we take advantage of additional information---knowing 
that a player wins $3-0$ is more information than knowing only that the player won. 
We fit separate models for women's matches and men's matches and thus estimate
home advantage separately. For all models, we use MCMC in Stan \citet{carpenter2017} 
for inference. Bayesian hierarchical models are discussed in depth in \citet{gelman2006,
gelman2013} including the use of Stan for inference. The model we use here is 
closely related to the model described in \citet{gelman_blog} for predicting the 
outcome of World Cup matches.

We estimate that home advantage adds $0.4$ games to a player's 
margin of victory in the men's game and $0.3$ games for women. 
The standard errors on both estimated effect sizes are roughly $0.1$. 
Our model estimates that for evenly matched players, if one has home 
advantage then his/her probability of winning increases to roughly 58\%/56\%.\footnote{A note of caution---mean margin of victory can be a counterintuitive due to the absence of the zero outcome (the absence of ties) and since our error model is symmetric and continuous, though the actual outcome is discrete.} 

\begin{figure}[t!]
\centering
\begin{subfigure}[t]{0.49\textwidth}
  \centering
  \includegraphics[width=\textwidth]{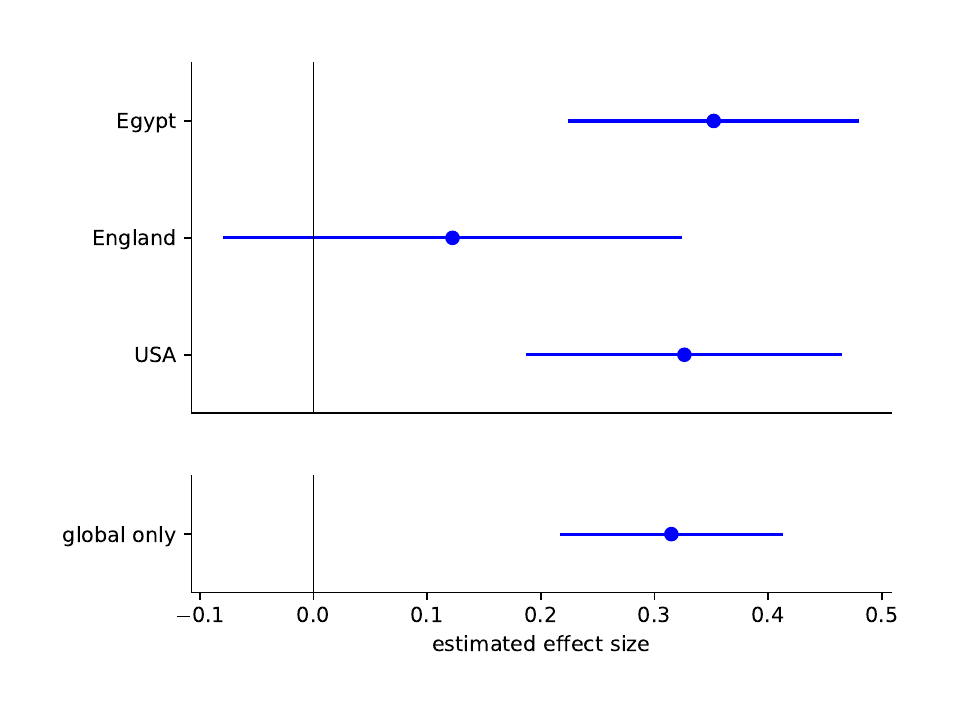}
  \caption{Women}
\end{subfigure}
%
\centering
\begin{subfigure}[t]{0.49\textwidth}
  \centering
  \includegraphics[width=\textwidth]{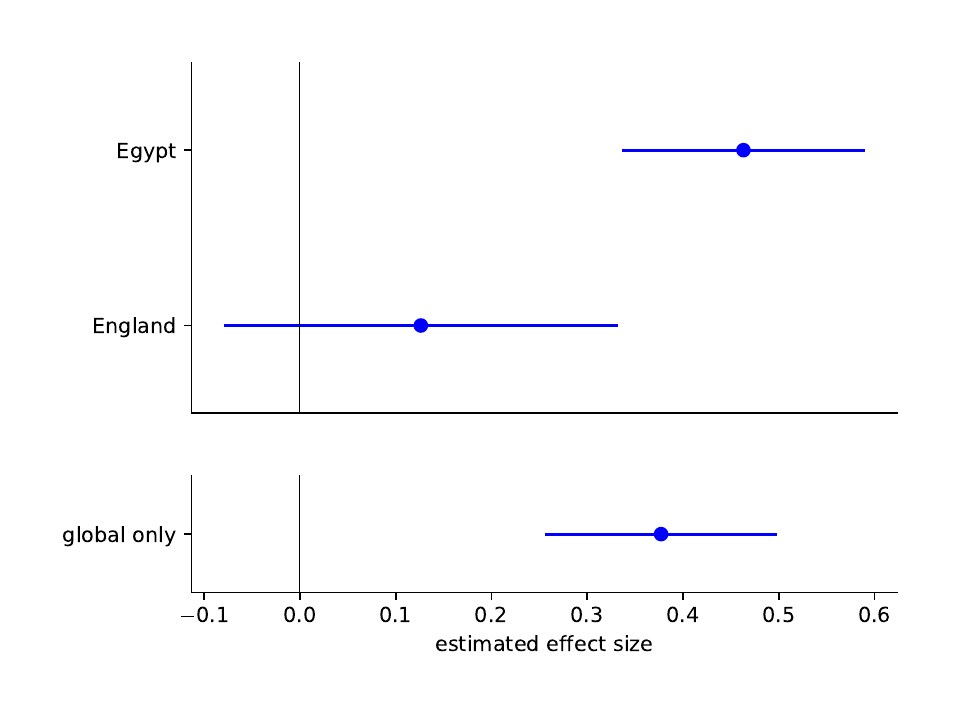}
\caption{Men}
\end{subfigure}
\caption{\em Estimated effect of home advantage using two models. On top axes, we show the estimated effect sizes of home advantage with model \eqref{model2}, which includes a  intercepts for Egypt, England, and U.S.. In the bottom set of axes, we measure global home advantage via model \eqref{model1}. Effect sizes for U.S. men are excluded due to small sample size. 
}
\label{fig:home_bars}
\end{figure}

We test whether the impact of home advantage is particularly large in Egypt, 
England, and the U.S., three countries that frequently host 
the PSA's most important tournaments. 
We use a model with a global intercept for home advantage, in addition to  
three country-specific intercepts for England, Egypt, and 
U.S. (see Section \ref{sec:methods} for details). 
The results are summarized in Figure \ref{fig:home_bars}. 
We estimate that the home advantage in Egypt is $0.45$ games for men 
and $0.35$ for women with standard errors of around $0.1$. We find that home 
advantage for women in the U.S. is similar to women in Egypt. 
Our model includes only PSA matches between players ranked in the top 30 in 
the world, thus we have more data (and less uncertainty in estimated effect sizes)
on home advantage in Egypt, the country with the most players in the top 30 for both men and women. 
Sample sizes are reported in Table \ref{t:matches_played}. 
Estimates of the effect of home advantage for American men and for English men 
and women have large standard errors. We don't have enough data to estimate
the effect with precision. 

\begin{figure}[h!]
\centering
\begin{subfigure}[t]{0.45\textwidth}
  \centering
  \includegraphics[width=\textwidth]{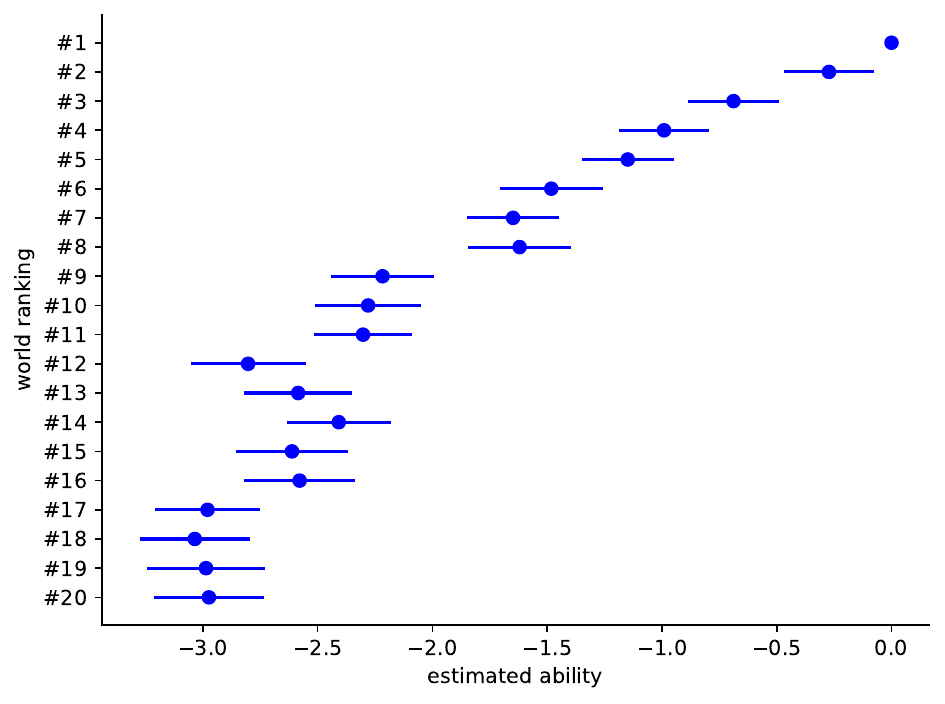}
  \caption{Women}
\end{subfigure}
\hspace{1em}
\centering
\begin{subfigure}[t]{0.45\textwidth}
  \centering
  \includegraphics[width=\textwidth]{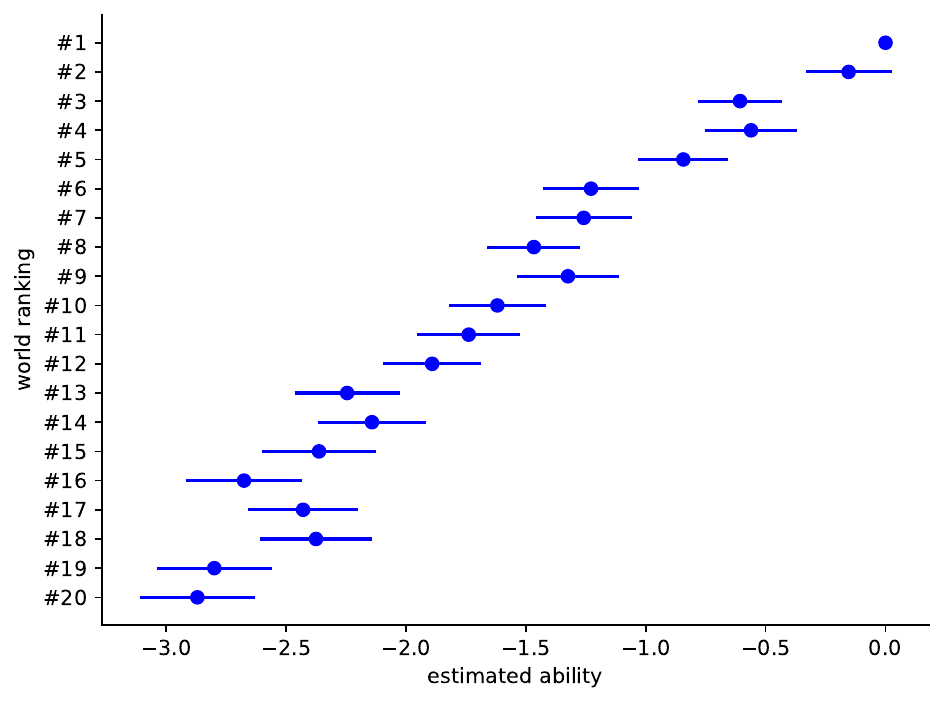}
\caption{Men}
\end{subfigure}
\caption{\em Estimated abilities $+/-$ one standard error for various world ranking levels.}
\label{fig:ranked_abilities}
\end{figure}

How big a difference does home advantage make if the estimated effect size is accurate? 
For the top 20 rankings in the world, we estimate that the average difference in ability
between one ranking and the next is 
roughly $0.15$ (see Figure \ref{fig:ranked_abilities}) for both men and women. 
A caveat---the difference 
in estimated abilities between successive players in the rankings depends 
on the ranking level. For instance, the difference between world \#2 and world \#3 is estimated
to be larger than the difference between \#3 and \#4. 
But on average, players that are next to each other in world rankings 
are pretty evenly matched---the mean difference in games won in a match between 
them is $0.15$. By comparison, the effect of home advantage is estimated to be 
roughly $0.3$ and $0.4$ for women and men. 
In Figures \ref{fig:matchups_f} and \ref{fig:matchups_m} we compare our model's
predictions (with and without home advantage) for to the actual 
actual result. We do this for two tournaments, 
one that took place in Egypt (El Gouna International) and another in England 
(British Open). 

Our analysis and model are simple and there are several improvements that
could be made. For example, some players might have the support of the crowd even 
if they're not in their home country. 
We don't take into account information about a match beyond the world ranking of each
player and whether one player has a home advantage. As a result, if certain players systematically have an 
ability that is not commensurate with their ranking, then errors can be introduced.
For example, we estimate that the world's \#4 ranked man is better than the 
\#3 . This is probably due to the fact that in our data set, the most games played by 
a men's \#4 ranking were played by Mostafa Asal who was suspended three 
times by the PSA during the time covered by our data and thus is probably under-ranked. 
In this study we do not look into the
potential causes of home advantage, we only estimate its effect on matches. 
There's a large literature on causes of home advantage and
various factors have been proposed including refereeing bias \citet{bbc_home_bias}, playing field conditions \citet{economist2012}, psychological impacts on players \citet{jamieson2010}, or many other explanations \citet{guardian2013}.

\section{Methods}\label{sec:methods}
We use a Bayesian hierarchical model to measure home advantage. 
We use the difference in the number of games won as the outcome. 
Since squash is usually played as a best of $5$ games, the outcome
takes values $y = -3,\ -2,\ -1,\ 1,\ 2,\ 3$. 
Our model takes into account only the world ranking of both players 
and whether one has a home advantage.
We use the linear model
\begin{linenomath}
\begin{align*}
y \sim \text{normal} (a_{\text{rank[player 1]}} -  a_{\text{rank[player 2]}} + h * b, \sigma_y)
\end{align*}
\end{linenomath}
where rank[player 1], rank[player 2] $\in \{1, 2, ...\}$ are the world rankings of 
player 1 and player 2 in a particular match. Here, $a_{i}$ corresponds to the ability 
of the player ranked \# $i$ in the world. The ability parameters $a_i$
depend only on the world ranking and not on which particular player occupied that 
world ranking slot at a particular time. 
We denote the home advantage effect by $h$ 
and the value $b \in \{-1, 0, 1\}$ encodes whether one of the two players in a
match has home advantage. We define home advantage by 
\begin{linenomath}
\begin{align}
b = 
\begin{cases} 
      1,  & \text{player $1$ is home and player $2$ is not} \\
      0, & \text{neither player $1$ nor $2$ are home or both $1$ and $2$ are home} \\
      -1,  & \text{player $1$ is home and player $2$ is not.}
   \end{cases}
   \nonumber
\end{align}
We assign the  prior on the home advantage intercept 
\begin{align}
h \sim \text{normal}(0, 0.5).
   \nonumber
\end{align}
\end{linenomath}
The mean of the likelihood function depends only on differences 
between the abilities of the two players, thus abilities are only
defined up to an additive constant. To account for this, we impose 
that the top-ranked 
player has an ability of zero.
We use a prior on all other abilities of 
\begin{linenomath}
\begin{align*}
    &a_{j} \sim \text{normal}(\beta (j - 1) + \gamma \sqrt{j - 1}, \sigma_{a}) \\
    &\beta \sim \text{normal}(0, 1) \\
    &\gamma \sim \text{normal}(0, 1)
\end{align*}
\end{linenomath}
for $j > 1$. 
The hierarchical prior on $a_j$ comes from the observation that differences between 
abilities are not linear in rank. For instance, the difference between the 
world's \#1 and \#11 player is likely much larger than the difference between 
the world's \#11 and \#21. Thus we include a linear and square root term in the 
mean of the prior on $a_j$. 
Lastly we give the standard deviations $\sigma_y, \sigma_{a}$ priors of 
\begin{equation}
\begin{aligned}
& \sigma_{y} \sim \text{normal}^+(0, 2)\\
& \sigma_{a} \sim \text{normal}^+(0, 2). 
\nonumber
\end{aligned}
\end{equation}
All together, our model is 
\begin{equation}\label{model1}
\begin{aligned}
& y \sim \text{normal} (a_{\text{rank[player 1]}} - a_{\text{rank[player 2]}} + h * b, \sigma_y) \\
    &a_{j} \sim \text{normal}(\beta (j - 1) + \gamma \sqrt{j - 1}, \sigma_{a}) \qquad \text{for $j=1,...,30$}\\
    &\beta \sim \text{normal}(0, 1) \\
    &\gamma \sim \text{normal}(0, 1) \\
& h \sim \text{normal}(0, 0.5) \\
& \sigma_y \sim \text{normal}^+(0, 2) \\
& \sigma_{a} \sim \text{normal}^+(0, 2). 
\end{aligned}
\end{equation}

In order to test whether home advantage is larger in Egypt, England, and the U.S., we use a similar
model with a global intercept for home advantage in addition to country-specific intercepts.
\begin{equation}\label{model2}
\begin{aligned}
& y \sim \text{normal} (a_{\text{rank[player 1]}} - a_{\text{rank[player 2]}} + h_{global} * b
+ h_{country} * b_{\text{country}}
, \sigma_y) \\
    &a_{j} \sim \text{normal}(\beta (j - 1) + \gamma \sqrt{j - 1}, \sigma_{a}) \qquad \text{for $j=1,...,30$} \\
    &\beta \sim \text{normal}(0, 2) \\
    &\gamma \sim \text{normal}(0, 2) \\
& h_{global} \sim \text{normal}(0, 0.5) \\
& h_{\text{country}} \sim \text{normal}(0, 0.2) \\
& \sigma_y \sim \text{normal}^+(0, 2) \\
& \sigma_{a} \sim \text{normal}^+(0, 2). 
\end{aligned}
\end{equation}

In Appendix \ref{a:stan_models} we include the Stan model implementations
of both model \eqref{model1} and \eqref{model2}.

\subsection*{Model deficiencies}
\begin{itemize}
\item
We only consider players' world rankings, not the players themselves. World rankings are, for several reasons, imperfect
proxies for abilities. Players can miss tournaments for various reasons including 
injury, suspension, and personal reasons. For example, Mostafa Asal
is likely regularly outperforming his world ranking due to suspensions. This is likely one reason why
our model estimates that the \#4 world ranked player is stronger than the third---the most games played
by a world ranked 4 player was played by Mostafa Asal. 

\item 
The model is not generative. The observation model is continuous but the outcome is discrete---it takes 
values in $\{-3, -2, -1, 1, 2, 3 \}$. However, we can round numbers to get predictions. 

\item 
If a country's players are underrated relative to their rankings, and those players also play their matches
predominantly at home, then this can result in an inflated home advantage for that country. 
\end{itemize}

\subsection*{Data choices}
We use PSA World Tour games from November 2018 to February 2024 including Bronze, Silver, Gold, and Platinum
levels. We also include PSA World Championships and the PSA World Tour Finals. 
We exclude matches that ended in one player retiring (including walkovers). 
For best of three matches, we multiply the outcome by $1.5$ to put the match 
on scale with the large majority of matches that are best of five. 
We did not remove matches that were played in front of small or no crowds 
when squash resumed after the COVID-19 shutdown.

\begin{table}
\centering
  \begin{subtable}[h]{0.45\linewidth}
    \centering
    \resizebox{0.7\linewidth}{!}{
    \begin{tabular}{lrr}

 &  women &  men \\
\midrule
        Egypt &    348 &  329 \\
      England &    128 &  252 \\
         U.S. &    419 &  426 \\
        other &    113 &  333 \\

\end{tabular}

    }
    \caption{\em total matches}
  \end{subtable}
    \begin{subtable}[h]{0.45\linewidth}
    \centering
    \resizebox{0.7\linewidth}{!}{
    \begin{tabular}{lrr}

 &  women &  men \\
\midrule
        Egypt &    181 &  177 \\
      England &     36 &   43 \\
         U.S. &    118 &    1 \\
        other &     16 &   22 \\

\end{tabular}

    }
    \caption{\em matches with home advantage}
  \end{subtable}
  \caption{\em Number of matches in data set by location of match and whether 
  one player had home advantage.}
  \label{t:matches_played}
\end{table}

\section{Results}
We implemented the global home advantage model (see \eqref{model1}) and the country-specific intercept model (see \eqref{model2}) in Stan and 
performed inference with Stan's MCMC sampler \citet{carpenter2017}. 
Our model estimates that home advantage adds 0.4 games to men's margin of victory and 0.3 games for women with standard errors on both effect sizes of $0.1$. Figure \ref{fig:home_bars} illustrates effect sizes and standard errors. To put these effects 
in context, When two men/women of equal ability play, 
a player with home advantage has a 58\%/56\% chance of winning.\footnote{We get these probabilities from our estimated global home advantage effect size ($\mu_{w} = 0.3$ for women and $\mu_m = 0.4$ for men) along with our estimate for the standard deviations of the likelihood function of $\sigma_w = 1.8, \sigma_m = 1.9$ for women and men. Denoting the cumulative distribution function of a normal by $\Phi$, we have $\Phi(\mu_w / \sigma_w) = 0.56$ and $\Phi(\mu_m / \sigma_m) = 0.58$.}

Using the country-specific model, we estimated effects of home advantage in Egypt, England, and the USA. We find that home advantage is especially large in Egypt, where we estimate the home advantage for men to be 0.45 games. For women, both Egypt and the U.S. show home advantage effects of roughly 0.3 games. England exhibits smaller home advantages for both men and women, though with larger standard errors due to fewer matches with home players, as detailed in Table \ref{t:matches_played}.

\begin{figure}[h!]
\centering
\begin{subfigure}{0.49\textwidth}
  \centering
  \includegraphics[width=1.0\textwidth]{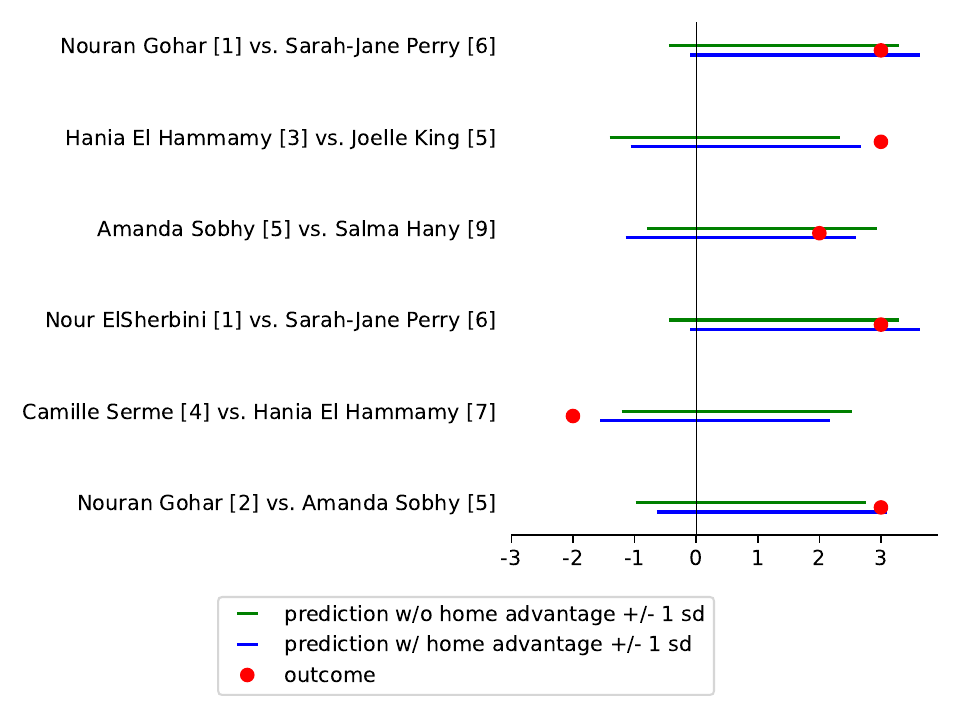}
  \caption{El Gouna (Egypt) women}
\end{subfigure}
\begin{subfigure}{0.49\textwidth}
  \centering
  \includegraphics[width=1.0\textwidth]{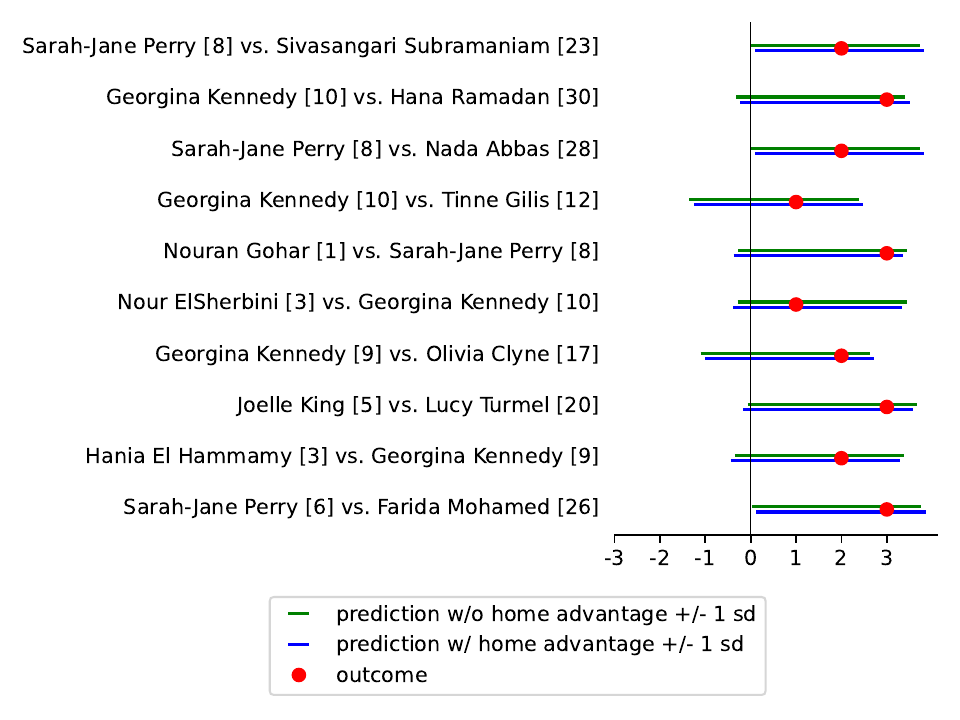}
\caption{British Open women}
\end{subfigure}
\caption{\em Recent women's British Open and El Gouna matchups 
where one of the players has home advantage. The outcome and predictions
are for the margin of victory of the player listed first (the higher ranked player) with 68\% predictive interval
from model \eqref{model2}. We plot the posterior credible intervals
for the outcome with home advantage (blue) and without (green).}
\label{fig:matchups_f}
\end{figure}

\begin{figure}[h!]
\centering
\begin{subfigure}{0.49\textwidth}
  \centering
  \includegraphics[width=1.0\textwidth]{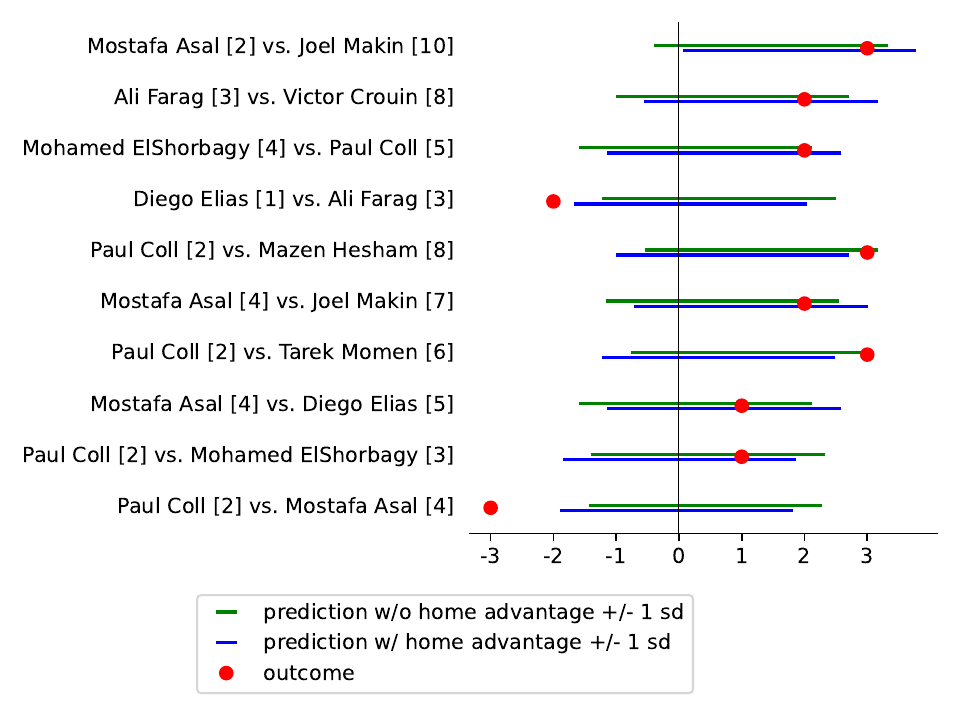}
  \caption{El Gouna (Egypt) men}
\end{subfigure}
\begin{subfigure}{0.49\textwidth}
  \centering
  \includegraphics[width=1.0\textwidth]{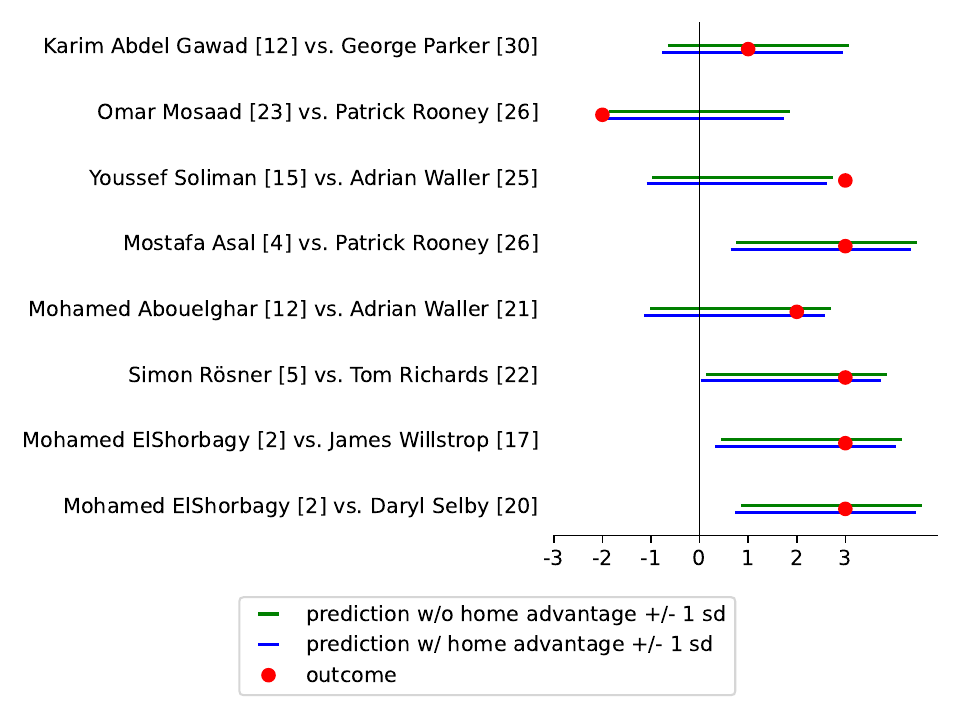}
\caption{British Open men}
\end{subfigure}
\caption{\em Recent men's British Open and El Gouna matchups 
where one of the players has home advantage. The outcome and predictions
are for the margin of victory of the player listed first (the higher ranked player) with 68\% predictive interval
from model \eqref{model2}. We plot the posterior credible intervals
for the outcome with home advantage (blue) and without (green).}
\label{fig:matchups_m}
\end{figure}

In Figures \ref{fig:matchups_f} and \ref{fig:matchups_m} we compare match outcomes to our predictions in cases where one player had home advantage. These figures demonstrate predicted and actual margin of victory for matches at two major tournaments, El Gouna International (Egypt) and the British Open.

\section{Discussion and conclusion}
We have estimated the effect of home advantage in professional squash using a Bayesian hierarchical model. We estimate that playing in one's home country would turn an otherwise even match to a win probability of 56\% (women) and 58\% (men) for the home player. The estimated effect is particularly strong in Egypt, though data limitations prevent precise estimation of country-specific effects elsewhere.

Several refinements could improve our analysis. Our model doesn't account for crowd support that players might receive even when not in their home country, nor does it incorporate match-specific features beyond rankings and home advantage. Additionally, players whose true ability differs systematically from their ranking (due to factors like suspensions or rapid improvement) may introduce bias in our estimates.

While this study quantifies home advantage, we don't investigate its underlying causes, which could include refereeing bias, familiarity with playing conditions, psychological factors, or other explanations proposed in sports science literature. Future work could explore these mechanisms in the context of professional squash.

\begin{figure}[h!]
\centering
\begin{subfigure}[t]{0.45\textwidth}
  \centering
  \includegraphics[width=\textwidth, height=3in]{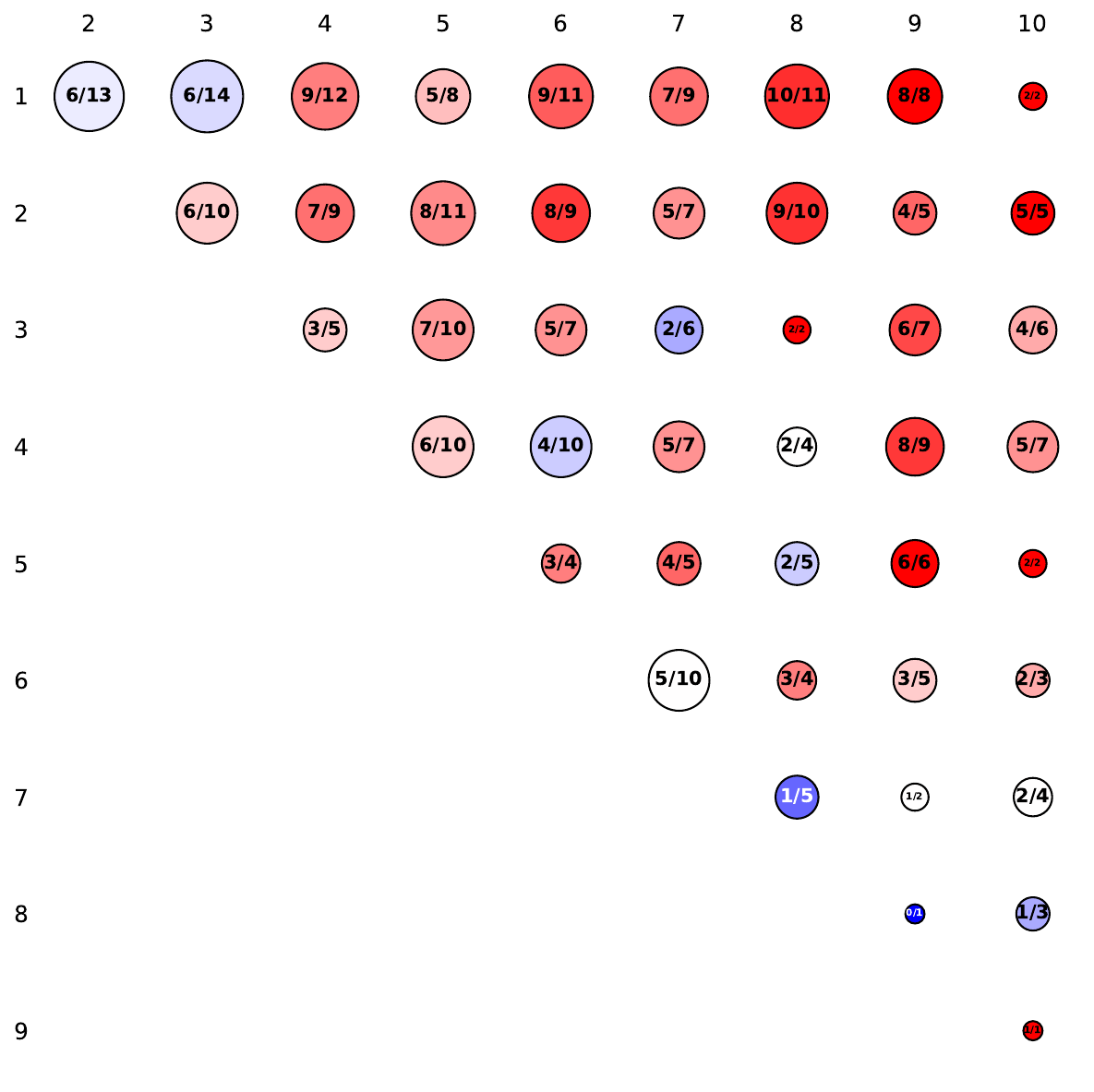}
  \caption{Women}
\end{subfigure}
\hspace{2em}
\centering
\begin{subfigure}[t]{0.45\textwidth}
  \centering
  \includegraphics[width=\textwidth, height=3in]{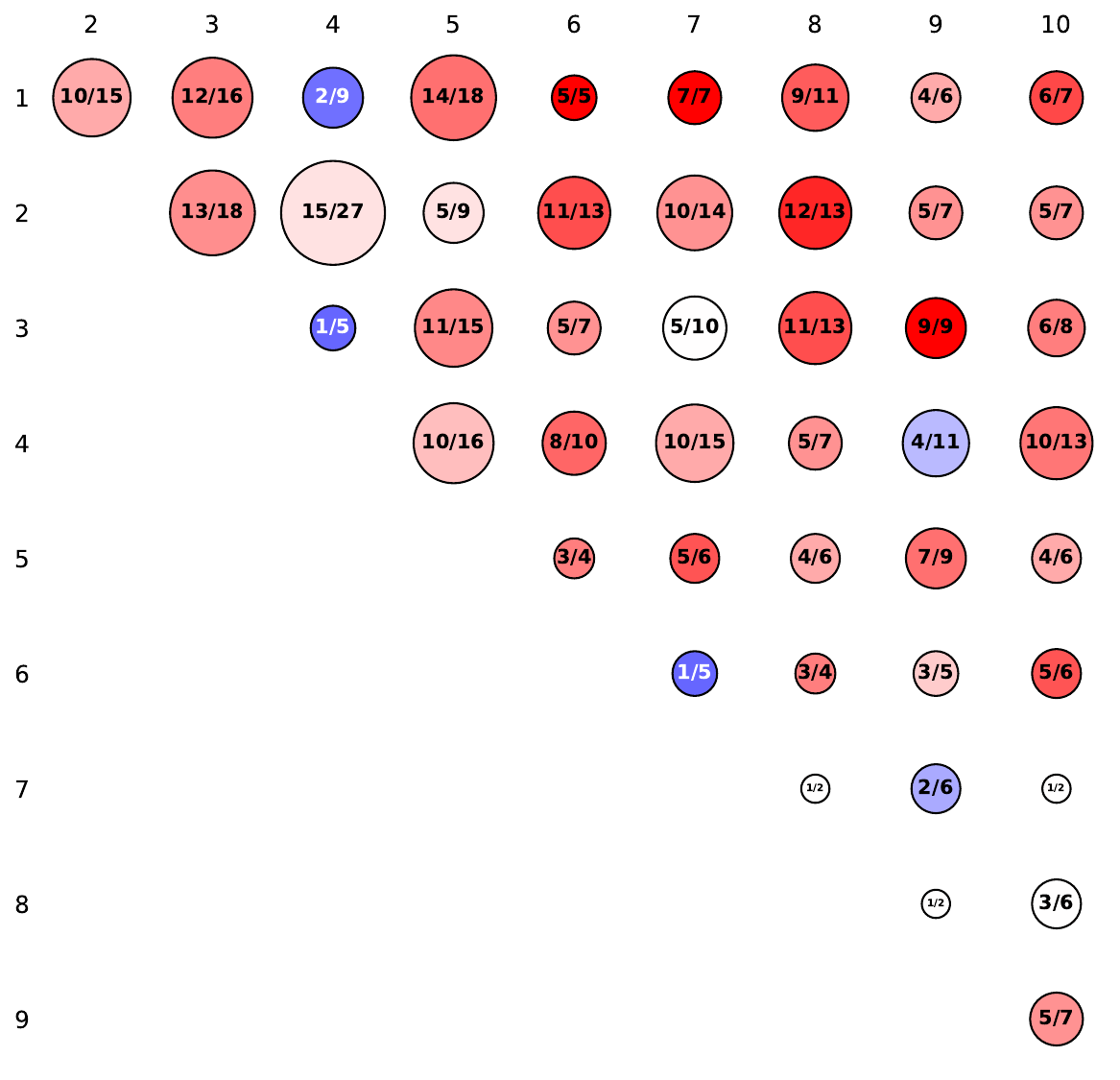}
\caption{Men}
\end{subfigure}
\caption{\em Head-to-head matchups of top ten world ranked players since 2018.
Fractions are the number of victories for the higher-ranked player out of the total number 
of matches. Winning records are in red and losing records are in blue. The size of the
dot is larger for when more matches in a particular matchup have been played.}
\label{fig:top10}
\end{figure}

\section*{Acknowledgments}
We thank Michael Abbotson, Jamie Harlington, and Andrew Gelman for useful discussions. We thank squashinfo.com and Howard Harding for providing PSA tour data.

\bibliographystyle{apalike}
\bibliography{refs}

\appendix
\section{Stan models}\label{a:stan_models}
\newpage
The Stan code we use for implementing model \eqref{model1} is:
\begin{verbatim}
data {
  int<lower=0> nplayers;
  int<lower=0> nmatches;
  array[nmatches] int<lower=0> player1;
  array[nmatches] int<lower=0> player2;
  array[nmatches] int home;
  array[nmatches] real<lower=1,upper=3> y;
}
parameters {
  array[nplayers - 1] real a0;
  real h;
  real<lower=0> sigma_y;
  real<lower=0> sigma_a;
  real beta;
  real gamma;
}
transformed parameters {
  array[1] real a1;
  a1[1] = 0;
  array[nplayers] real a = append_array(a1, a0);
}
model {
  for (i in 1:nmatches) {
    y[i] ~ normal(a[player1[i]] - a[player2[i]] + h * home[i], sigma_y);
  }
  for (i in 2:nplayers) {
    a[i] ~ normal(beta * (i - 1) + gamma * sqrt(i - 1), sigma_a);
  }
  beta ~ normal(0.0, 2.0);
  gamma ~ normal(0, 2.0);
  h ~ normal(0.0, 0.5);
  sigma_y ~ normal(0, 2);
  sigma_a ~ normal(0, 2);
}
\end{verbatim}

The Stan code we use for implementing model \eqref{model2} is:
\newpage 
\begin{verbatim}
data {
  int<lower=0> nplayers;
  int<lower=0> nmatches;
  array[nmatches] int<lower=0> player1;
  array[nmatches] int<lower=0> player2;
  array[nmatches] int home;
  array[nmatches] int home_egy;
  array[nmatches] int home_eng;
  array[nmatches] int home_usa;
  array[nmatches] real<lower=1,upper=3> y;
}
parameters {
  array[nplayers - 1] real a0;
  real h;
  real egy;
  real eng;
  real usa;
  real<lower=0> sigma_y;
  real<lower=0> sigma_a;
  real beta;
  real gamma;
}
transformed parameters {
  array[1] real a1;
  a1[1] = 0;
  array[nplayers] real a = append_array(a1, a0);
}
model {
  for (i in 1:nmatches) {
    y[i] ~ normal(a[player1[i]] - a[player2[i]] + h * home[i]
    + egy * home_egy[i]
    + eng * home_eng[i]
    + usa * home_usa[i], sigma_y);
  }
  for (i in 2:nplayers) {
    a[i] ~ normal(beta * (i - 1) + gamma * sqrt(i - 1), sigma_a);
  }
  beta ~ normal(0.0, 2.0);
  gamma ~ normal(0, 2.0);
  h ~ normal(0.0, 0.5);
  egy ~ normal(0, 0.2);
  eng ~ normal(0, 0.2);
  usa ~ normal(0, 0.2);
  sigma_y ~ normal(0, 2);
  sigma_a ~ normal(0, 2);
}
\end{verbatim}

\end{document}